\documentclass[useAMS, apjl]{emulateapj}
\usepackage{graphicx, hyperref}
\usepackage{natbib}
\usepackage{amsmath,amssymb}
\usepackage{aas_macros}

\newcommand{\Teff}{$T_{\rm eff}$}
\long\def\symbolfootnote[#1]#2{\begingroup%
\def\thefootnote{\fnsymbol{footnote}}\footnote[#1]{#2}\endgroup}

\begin{document}

\shorttitle{A Li-rich giant in NGC 362}
\shortauthors{D'ORAZI ET AL.}

\title{On the serendipitous discovery of a Li-rich giant in the globular cluster NGC 362\footnotemark[1]}
\footnotetext[1]{Based on observations taken with ESO telescopes under program 094.D-0276(A)}

\author{ Valentina D'Orazi\altaffilmark{2,3,4}, Raffaele G. Gratton \altaffilmark{2}, George C. Angelou\altaffilmark{5,6}, Angela Bragaglia\altaffilmark{7}, Eugenio Carretta\altaffilmark{7}, 
John C. Lattanzio \altaffilmark{4}, Sara Lucatello\altaffilmark{2}, Yazan Momany\altaffilmark{2,8}, and Antonio Sollima \altaffilmark{7}}

\altaffiltext{2}{INAF - Osservatorio Astronomico di Padova, Vicolo dell'Osservatorio 5, 35122, Padova, Italy. Email: \email{valentina.dorazi@oapd.inaf.it}}
\altaffiltext{3}{Department of Physics and Astronomy, Macquarie University, Sydney, NSW 2109, Australia.} 
\altaffiltext{4}{Monash Centre for Astrophysics (MoCA), Monash University,  Melbourne,  VIC 3800,  Australia.}
\altaffiltext{5}{Max Planck Institut f\"{u}r Sonnensystemforschung, Justus-von-Liebig-Weg 3, 37077 G\"{o}ttingen, Germany} 
\altaffiltext{6}{Stellar Astrophysics Centre, Department of Physics and Astronomy, Aarhus University, Ny Munkegade 120, 8000 Aarhus C, Denmark}
\altaffiltext{7}{INAF - Osservatorio Astronomico di Bologna, via Ranzani 1, 40127, Bologna, Italy }
\altaffiltext{8}{European Southern Observatory, Alonso de Cordova 3107, Vitacura, Santiago, Chile}

\vspace{1em}

\begin{abstract}
We have serendipitously identified the first lithium-rich giant star located close to the red giant branch bump in a globular cluster.
Through intermediate-resolution FLAMES spectra we derived a lithium abundance of  A(Li)=2.55 (assuming local thermodynamical equilibrium), which is  extremely high considering the
star's evolutionary stage. Kinematic and photometric analysis confirm the object as a member of the globular cluster NGC 362.
This is the fourth Li-rich giant discovered in a globular cluster but the only one known to exist at a luminosity close to the bump magnitude. The three previous detections are clearly more evolved, located close to, or beyond the tip of their red giant branch. Our observations are able to discard the  accretion of planets/brown dwarfs, as well as an enhanced mass-loss mechanism as a formation channel for this rare object.
Whilst the star sits just above the cluster bump luminosity, its temperature places it towards the blue side of the giant branch in the colour-magnitude diagram. We require further dedicated observations to unambiguously identify the star as a red giant: we are currently unable to confirm whether Li production has occurred at the bump of the luminosity function or if the  star is on the pre zero-age horizontal branch. 
The latter scenario provides the opportunity for the star to have synthesised Li rapidly during the core helium flash or gradually during its red giant branch ascent via some extra mixing process.
\end{abstract}


\section{Introduction} \label{sec:intro}

Lithium is one of the most fragile elements, promptly consumed by
(p,$\alpha$) reactions when the temperature in stellar interiors
reaches T$\approx$2.5$\times10^6$ K. According to standard stellar
evolution theory, when a star leaves the main-sequence and experiences the first dredge-up (\citealt{iben67}), the convective envelope penetrates into hot internal layers, mixing the outer zones with matter that has undergone partial H-burning.
The main surface abundance changes are increases in $^4$He, $^{14}$N
and $^{13}$C, and a decrease in the $^{12}$C abundance
(e.g., \citealt{karakas10}). The initial Li content is reduced by
factors 15$-$20 in Population~{\sc ii} stars,
reaching typical values of A(Li)~$\leq$1.5 dex. Observations of
Li abundances in low-mass red giant-branch (RGB) stars have indeed
provided strong and unambiguous confirmation to these theoretical
predictions (e.g., \citealt{lambert80}; \citealt{gratton00}). Yet,
approximately between 1$-$2 \% of K giants have been found to exhibit
significantly higher Li abundances and they are thus referred to as
Li-rich giants (\citealt{brown89};
\citealt{monaco11}; \citealt{kumar11}; \citealt{lebzelter12} and references therein).
Interestingly, this fraction might be even lower, that is
0.56\%$\pm$0.16\%, as recently emphasised by \cite{zhang15}.
A clear explanation for this rare phenomenon is still wanting and
different mechanisms have been suggested. They envisage contamination from
ejecta of nearby novae
(\citealt{martin94}), accretion of planets and/or brown dwarfs
(\citealt{sl99}),
as well as freshly synthesised Li in the stars themselves (see e.g.,
\citealt{sb99}).
The physical process responsible for Li production is the well known
\citeauthor{cf71} (\citeyear{cf71})
mechanism: the fusion sequence is
$^3$He($\alpha$,$\gamma$)$^7$Be($e$,$\nu$)$^7$Li, with the requirement
that the astrated $^7$Be is quickly transported to the cooler envelope
so that the eventual $^7$Li produced will not be destroyed by proton
captures.
\cite{charb00} proposed two distinct
episodes of Li production depending upon the mass of the star. In low-mass RGB stars, with degenerate helium cores, Li production may be associated with the bump in the luminosity function; alternatively, intermediate-mass stars can
synthesise Li once the core-He burning has ceased and they ascend
the early asymptotic giant branch (AGB, see that paper for details).

The preliminary observational detections of  Li-rich giants seemed to suggest that these stars occupy only specific regions of the Hertzsprung-Russel diagram (such as the RGB bump, tip and/or the red clump). However,  several studies (e.g., \citealt{monaco11}; \citealt{lebzelter12}) have recently identified  Li-rich stars at magnitudes that span the red giant branch.
Their results indicate that the formation of Li-rich giants might not be limited to the aforementioned specific evolutionary events.

Given how rare these stars are, large surveys are required (and underway) to increase the number of observed Li-rich objects.
More detections are necessary to improve our understanding of the physical phenomena involved and provide empirical constraints on their formation. 
Many stellar environments,  within and beyond the Milky Way,  have been scrutinised for the presence of Li-rich giants
(thin/thick disk, bulge and halo in the Milky Way, dwarf
galaxies, open and globular clusters -GCs, e.g.,
\citealt{monaco11}; \citealt{lebzelter12}; \citealt{ms13};
\citealt{kirby12}; \citealt{monaco14}; \citealt{pilachowski00}).
Clusters (either open or globular) offer an obvious advantage in that they host
mono-metallic, almost coeval stellar populations and hence represent an
excellent
target for this kind of investigation: their masses and luminosities
are also much more accurately determined with respect to field stars.
To date, only three Li-rich giants have been reported in GCs: V42 in M5
(\citealt{carney98}),
V2 in NGC 362 (\citealt{smith99}), and
IV-101 in M3 (\citealt{kraft99}). All of them are located in the upper
part of the RGB: V2 and IV-101 lie at the tip of the RGB, while V42 is
actually a
W Vir post-AGB star.
We present in this {\it Letter} the serendipitous discovery of a Li-rich
giant close the RGB bump in the globular cluster NGC~362, a relatively metal-rich cluster ([Fe/H]=$-$1.26 dex, 2011 update of the Harris' catalogue), characterised by chaotic and unusual orbital parameters (\citealt{dinescu99}). 
This is the
first time that a Li-rich giant is detected at this evolutionary stage
in a GC.


\section{Observations and analysis}\label{sec:obs}
Observations were conducted in visitor mode with FLAMES (\citealt{pasquini02}) mounted at ESO VLT on 11, 12, and 13 December 2014, 
under program 094.D-0276(A). Spectra were collected with the scientific aim of inferring Li and Al abundances for large samples of RGB stars in several GCs, as part of our dedicated survey (see \citealt{dorazi14}, hereinafter Paper~{\sc i}). 
We utilised the HR15N setup that provides a nominal resolution of R=17,000 and a spectral coverage from 6470 \AA~to 6790 \AA, allowing the simultaneous inclusion of the strong doublets for Li~{\sc i} (6708 \AA) and Al~{\sc i} (6696/6698 \AA). 
The data were reduced using the dedicated ESO software (version 2.12.2, available at \url{http://www.eso.org/sci/software/pipelines/})
and resulting in bias-subtracted, flat-fielded, wavelength-calibrated, one-dimensional spectra. Sky subtraction and rest-frame shift were then carried out within IRAF\footnote{IRAF is the Image Reduction and Analysis Facility, a general purpose software system for the reduction and analysis of astronomical data. IRAF is written and supported by National Optical Astronomy
Observatories (NOAO) in Tucson, Arizona.}.
As with Paper~{\sc i}, stellar parameters were derived in the following manner: effective temperatures ($T_{\rm eff}$) are inferred from $V$ 
 and K$_s$ photometry (from \citealt{momany04} and \citealt{skru06}, respectively) and the calibration by \cite{alonso99}. 
 The reddening
and metallicity values are adopted from the \cite{harris96} catalogue, that is E($B-V$)=0.05 and [Fe/H]=$-$1.26 dex. 
Surface gravities were computed from the standard formula relating luminosities, temperatures and masses\footnote{$$\log\frac{g}{g_\odot}=\log\frac{M}{M_\odot}-\log\frac{L}{L_\odot}+4\log\frac{T_{\rm eff}}{T_{\rm eff,\odot}}$$}, 
adopting M=0.85 M$_\odot$ and bolometric solar magnitude of M$_{bol, \odot} = 4.75$. Microturbulence velocities ($\xi$) are from the relation by \cite{gratton96}. Abundances were obtained via spectral synthesis for both Li and Al using the local thermodynamical equilibrium (LTE) code {\sc moog} (\citealt{sneden73}, 2014 version); interpolated Kurucz model atmospheres based on the ATLAS9
code with $\alpha$-enhancement and no convective overshooting (\citealt{ck04})
were used throughout this study\footnote{\url{http://kurucz.harvard.edu/grids.html}}.

We adopted the same line list as in Paper~{\sc i}, performing the comparison between synthetic and observed spectra between 6696 and 6720 \AA~(exploiting the Ca~{\sc i} line to evaluate the spectral broadening). 
Errors related to best fit determination and uncertainties in stellar parameters were consistently calculated as for Paper~{\sc i} and then added in quadrature, given their independence. The total internal random error results in 0.09 dex and 0.12 dex for Li and Al, respectively.  
Finally, non-LTE corrections were applied to Li abundances, following prescriptions by \cite{lind09}.


\section{Results and Discussion}\label{sec:results}

Stellar parameters, kinematics and abundances for Li and Al are reported in Table~\ref{t:param} for our target star.
We inferred a Li abundance of 
A(Li)=2.55$\pm$0.09 (0.13 dex lower when departures from LTE are taken into account) which is significantly higher than 
our determined average for the cluster (A(Li)=0.90$\pm$0.02 rms=0.15, 66 RGB stars; D'Orazi et al. 2015, in preparation). 
Thus, star \#15370 joins the small family of Li-rich giants in GCs whose other members include a star in M3, M5 as well 
as a neighbour also in NGC~362. Star \#15370 distinguishes itself because it is the first Li-rich giant discovered in a GC 
near the luminosity of the RGB bump (roughly 0.12 magnitudes brighter than V$_{bump}$=15.40 from \citealt{nataf13}). 
Its siblings all sit close to the tip of their giant branches.
\begin{figure}
\centering
\includegraphics[width=0.9\columnwidth]{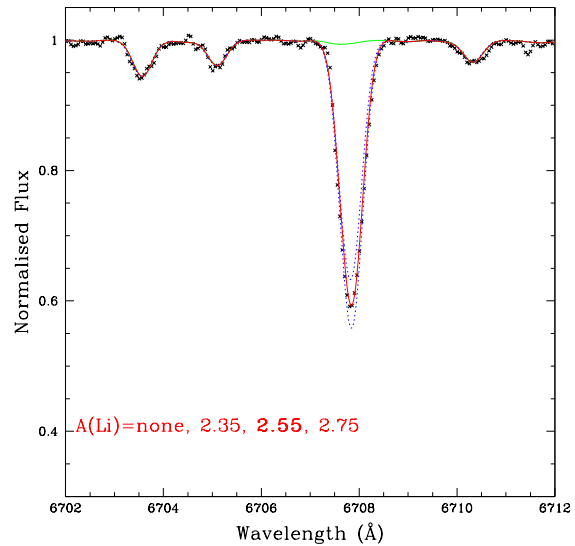}
\caption{Spectral synthesis of the Li~{\sc i} line at 6707.78 \AA~for our target star}
\label{f:synt}
\end{figure}
There is little doubt that the star is Li rich: the intrinsic strength of the 6708 \AA~doublet is evident from Figure~\ref{f:synt}, where we show the observed spectrum along with synthetic profiles calculated for different Li abundances. The equivalent width (EW), measured using the IRAF task $splot$ and a gaussian profile, is EW=250 m\AA.
Unfortunately, our spectral coverage does not include the other Li~{\sc i} feature at 6104 \AA, 
thus we cannot utilise this line as a secondary indicator. 

\begin{figure}
\centering
\includegraphics[width=0.9\columnwidth]{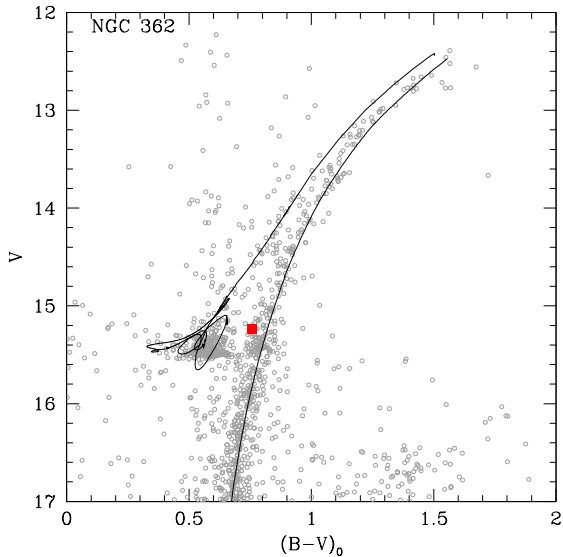}
\caption{Colour-magnitude diagram for NGC~362 (Momany et al. 2004). The target star is emboldened.}
\label{f:cmd}
\end{figure}

Kinematic information confirms cluster membership. The star's radial velocity V$_{rad}$= 222.68$\pm$1.02 km~s$^{-1}$ 
is consistent with the cluster average we derived from our sample of 71 RGB stars, i.e., V$_{rad}$= 222.78$\pm$0.62 km~s$^{-1}$.
This value agrees very well with V$_{rad}$=223.5$\pm$0.5 km~s$^{-1}$ as given by \cite{harris96}.
Moreover, the position on the colour-magnitude diagram (Figure~\ref{f:cmd}) 
as well as the spectral comparison with another cluster member with almost identical stellar parameters 
(see Figure 3: namely star \#2213, $T_{\rm eff}$=4821K, log$g$=2.19 cm~s$^{-2}$, $\xi$=1.51 km~s$^{-1}$) 
further advocate that this Li-rich giant belongs to NGC~362.

We find an Al abundance of [Al/Fe]=0.08 dex, which according to the definition by e.g., \cite{carretta09a} 
and Paper~{\sc i} suggests that our 
Li-rich giant is a member of the first generation of the cluster population\footnote{A discussion on the multiple population phenomenon in GCs is not the purpose of the current manuscript; we refer the reader to recent reviews by \cite{gratton12} and \cite{piotto12} 
for spectroscopic and photometric perspectives, respectively.} . 
A similar result was found by \cite{smith99} who derived high O and low Na and Al abundances for their Li-rich giant V2. 
Both IV-101 and V42 seem to show a (modest) Na enrichment and probably belong to the second generation of their respective clusters. 
Given the quite limited sample we are currently dealing with, statistics on the occurrence of Li-rich stars 
among the different stellar populations in GCs is not meaningful. We hope that the dedicated surveys will lead to a higher number of detections and help with such analyses.

One possible explanation for the formation of Li-rich K giants is by accretion of planets and/or brown dwarfs (\citealt{sl99}).
Given the high Li abundance of A(Li~{\sc i})$\approx$2.5 dex, this scenario seems unlikely for our star because it would require the ingestion of a very massive companion. Giants during this phase of evolution typically have a convective envelope of 
$\sim$0.5 M$_\odot$. To account for the observed Li abundance, star \#15370 needs to have engulfed a companion with mass of at least $\approx$1.0 M$_\odot$ 
(using Equation 2 given by \citealt{sl99} and adopting a Li content for the donor star of 2.55 dex, which is already an unusually high value with respect to the Spite plateau).
 Such a big accretion event should also result in increased rotational velocity and/or enhanced level of chromospheric activity (see e.g., \citealt{hurley02}; \citealt{drake02}), which is not evident from our spectra (at least within the limits imposed by the quality of our data).

 \citeauthor{delareza96} (\citeyear{delareza96},\citeyear{delareza97}) 
proposed a model connecting the high Li abundances observed in K giants to the evolution of circumstellar shells. In this framework, the internal mechanism responsible for the Li enrichment will initiate a prompt mass-loss event. 
Empirical evidence for this paradigm has not been forthcoming (e.g., \citealt{fekel98}).
We inspected our spectra around the H$\alpha$ feature, looking for possible blue-shifted asymmetry as a hint of enhanced (gas) mass-loss. 
Within the limit of our S/N and resolution, we did not discover any asymmetry in the H$\alpha$ profile of star \#15370 compared 
to other Li-normal giants with similar atmospheric parameters. In this respect, we confirm results published by \cite{monaco11} and \cite{lebzelter12}.

The literature offers two alternative scenarios that are consistent with our object:
\begin{itemize}
\item[$(i)$] Star \#15370 belongs to the class of Li-rich giants described by \citet{charb00} that are synthesing Li at the beginning of the RGB bump phase. After the hydrogen-burning shell removes the composition discontinuity left behind by first dredge-up, some extra-mixing process facilitates Li production via the Cameron \& Fowler mechanism. Subsequently, once the mixing deepens enough to convert additional $^{12}$C to $^{13}$C, the surface material is exposed to temperatures that are too high  and the fresh Li is quickly destroyed. Thus, the Li-rich phase is swift and a star will not retain its peak Li abundance once the $^{12}$C/$^{13}$C drops below the standard value (\citealt{charb00}). 
Parametrised analyses, unsurprisingly, require that these stars have unusually large diffusion coefficients to allow for the rapid transport of 
$^7$Be (\citealt{sb99}; \citealt{denher04}).

\item[$(ii)$] A second possibility is that the production of Li might have occurred during the helium flash (\citealt{kumar11}) or gradually as the star approached the RGB tip due to some kind of extra mixing (\citealt{lattanzio15}). Asteroseismic analysis of KIC 5000307 by \cite{silva14} has identified this Li-rich field star as clearly undergoing core helium burning. An analogous scenario for star \#15370 would require that it is actually a pre zero age horizontal branch (ZAHB) star. A close inspection of the CMD indicates that the star is located on the blue side of the RGB and its position is not inconsistent with theoretical predictions of pre-ZAHB evolution published by \cite{silva08}.  

\end{itemize} 

As a further check we have run evolutionary models using MONSTAR (Monash Version of the Mt. Stromlo evolution code, see \citealt{angelou12}).
For our model we assumed a metallicity of [Fe/H]=$-$1.26 dex, an $\alpha$-element enhancement of [$\alpha$/Fe]=0.2, 
and initial mass of 0.81 M$_\odot$ corresponding to a main-sequence turn off age of 11.5 Gyr (\citealt{dotter11}). 
As can be see from Figure~\ref{f:cmd} the star's location is best explained by RGB evolutionary tracks.
We note that although the star is (slightly) on the blue side of the giant branch it is still somewhat removed from the cluster HB.
However, in order to unequivocally determine the evolutionary phase, we plan to acquire high-resolution (R$\approx$40,000), high S/N spectra ($\geq$100) of our target star along with one of its siblings to derive their carbon isotopic ratios $^{12}$C/$^{13}$C.  
In principle, differences in Fe~{\sc ii} lines would also provide us with evidence of a slightly different gravity for \#15370 -- an indication that the star experienced mass loss at the tip of the RGB branch (a difference in mass of approximately 0.2 M$_\odot$ implies a difference in gravity of about 0.10 dex).

\begin{figure*}
\centering
\includegraphics[width=1.7\columnwidth]{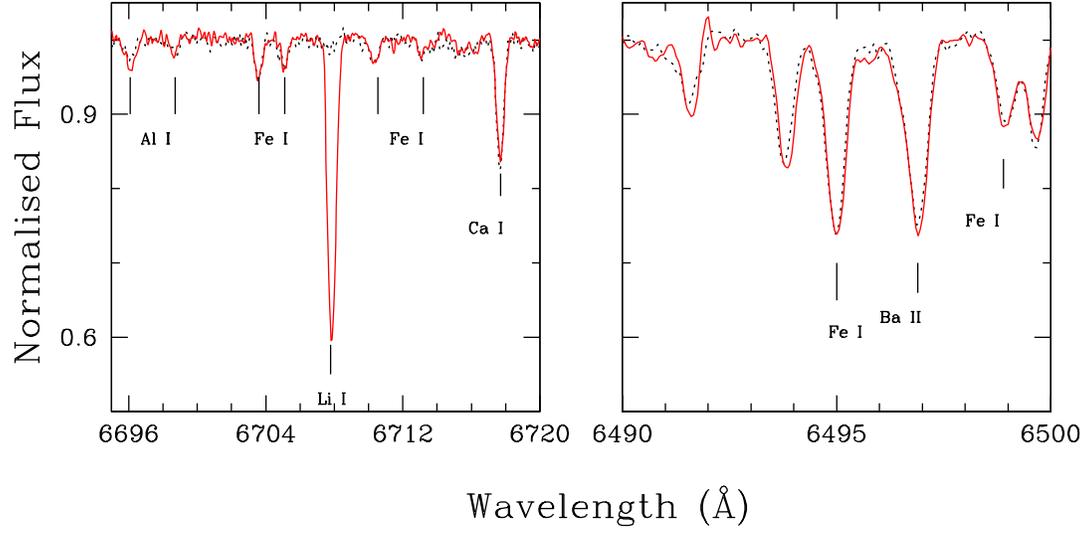}
\caption{Spectral comparison for stars \#15370 and \#2213.}
\label{fig:comparison}
\end{figure*}
\begin{table*}
\centering
\caption{Stellar parameters, kinematics and chemical information for star \#15370}\label{t:param}
\hspace*{-0.9cm}
\tabcolsep=0.10cm
\small
\begin{tabular}{lccccccccccccc}
  \hline \hline
Star ID & RA & DEC & $B$   & $V$   &  K$_s$ & V$_{\rm rad}$  &  S/N   & \Teff &  $\log \ g$     & $\xi$ & A(Li)$_{\rm LTE}$ & A(Li)$_{\rm NLTE}$ & [Al/Fe] \\
        &  (hh:mm:ss) & ($^{\circ}$:':'') & (mag) & (mag) &  (mag) &  (km s$^{-1}$) & (@6708\AA) & (K) &  (cm s$^{-2}$) & (km s$^{-1}$) & dex & dex & dex\\
  \hline
15370  &  01:03:43.9  &  -70:52:20.9   & 16.052 & 15.233 & 12.944 & 233.47 & 190 & 4812 & 2.18 & 1.52 & 2.55  & 2.42 & 0.08  \\
\hline
\end{tabular}
\label{tab:param}
\end{table*}     

%
%


\acknowledgments{This work made extensive use of the SIMBAD, Vizier, and NASA ADS databases. We acknowledge partial support by 
the Australian Research Council (ARC), PRIN INAF 2011 ``Multiple populations in globular clusters: their role in the Galaxy
assembly'' (PI E. Carretta) and PRIN MIUR 2010-2011 ``The Chemical and Dynamical Evolution of the Milky Way and Local
Group Galaxies” (prot. 2010LY5N2T, PI F. Matteucci). 
The research leading to the presented results has received funding from the European Research Council under the European Community's Seventh Framework Programme (FP7/2007-2013) / ERC grant agreement no 338251 (StellarAges). We thank the referee, Piercarlo Bonifacio, for very helpful and valuable comments and suggestions.}

\bibliographystyle{apj}

\begin{thebibliography}{47}
\expandafter\ifx\csname natexlab\endcsname\relax\def\natexlab#1{#1}\fi

\bibitem[{{Alonso} {et~al.}(1999){Alonso}, {Arribas}, \&
  {Mart{\'{\i}}nez-Roger}}]{alonso99}
{Alonso}, A., {Arribas}, S., \& {Mart{\'{\i}}nez-Roger}, C. 1999, \aaps, 140,
  261

\bibitem[{{Angelou} {et~al.}(2012){Angelou}, {Stancliffe}, {Church},
  {Lattanzio}, \& {Smith}}]{angelou12}
{Angelou}, G.~C., {Stancliffe}, R.~J., {Church}, R.~P., {Lattanzio}, J.~C., \&
  {Smith}, G.~H. 2012, \apj, 749, 128

\bibitem[{{Brown} {et~al.}(1989){Brown}, {Sneden}, {Lambert}, \&
  {Dutchover}}]{brown89}
{Brown}, J.~A., {Sneden}, C., {Lambert}, D.~L., \& {Dutchover}, Jr., E. 1989,
  \apjs, 71, 293

\bibitem[{{Cameron} \& {Fowler}(1971)}]{cf71}
{Cameron}, A.~G.~W., \& {Fowler}, W.~A. 1971, \apj, 164, 111

\bibitem[{{Carney} {et~al.}(1998){Carney}, {Fry}, \& {Gonzalez}}]{carney98}
{Carney}, B.~W., {Fry}, A.~M., \& {Gonzalez}, G. 1998, \aj, 116, 2984

\bibitem[{{Carretta} {et~al.}(2009){Carretta}, {Bragaglia}, {Gratton},
  {Lucatello}, {Catanzaro}, {Leone}, {Bellazzini}, {Claudi}, {D'Orazi},
  {Momany}, {Ortolani}, {Pancino}, {Piotto}, {Recio-Blanco}, \&
  {Sabbi}}]{carretta09a}
{Carretta}, E., {et~al.} 2009, \aap, 505, 117

\bibitem[{{Castelli} \& {Kurucz}(2004)}]{ck04}
{Castelli}, F., \& {Kurucz}, R.~L. 2004, ArXiv Astrophysics e-prints

\bibitem[{{Charbonnel} \& {Balachandran}(2000)}]{charb00}
{Charbonnel}, C., \& {Balachandran}, S.~C. 2000, \aap, 359, 563

\bibitem[{{de La Reza} {et~al.}(1996){de La Reza}, {Drake}, \& {da
  Silva}}]{delareza96}
{de La Reza}, R., {Drake}, N.~A., \& {da Silva}, L. 1996, \apjl, 456, L115

\bibitem[{{de La Reza} {et~al.}(1997){de La Reza}, {Drake}, \& {da
  Silva}}]{delareza97}
---. 1997, \apss, 255, 251

\bibitem[{{Denissenkov} \& {Herwig}(2004)}]{denher04}
{Denissenkov}, P.~A., \& {Herwig}, F. 2004, \apj, 612, 1081

\bibitem[{{Dinescu} {et~al.}(1999){Dinescu}, {Girard}, \& {van
  Altena}}]{dinescu99}
{Dinescu}, D.~I., {Girard}, T.~M., \& {van Altena}, W.~F. 1999, \aj, 117, 1792

\bibitem[{{D'Orazi} {et~al.}(2014){D'Orazi}, {Angelou}, {Gratton}, {Lattanzio},
  {Bragaglia}, {Carretta}, {Lucatello}, \& {Momany}}]{dorazi14}
{D'Orazi}, V., {Angelou}, G.~C., {Gratton}, R.~G., {Lattanzio}, J.~C.,
  {Bragaglia}, A., {Carretta}, E., {Lucatello}, S., \& {Momany}, Y. 2014, \apj,
  791, 39

\bibitem[{{Dotter} {et~al.}(2011){Dotter}, {Sarajedini}, \&
  {Anderson}}]{dotter11}
{Dotter}, A., {Sarajedini}, A., \& {Anderson}, J. 2011, \apj, 738, 74

\bibitem[{{Drake} {et~al.}(2002){Drake}, {de la Reza}, {da Silva}, \&
  {Lambert}}]{drake02}
{Drake}, N.~A., {de la Reza}, R., {da Silva}, L., \& {Lambert}, D.~L. 2002,
  \aj, 123, 2703

\bibitem[{{Fekel} \& {Watson}(1998)}]{fekel98}
{Fekel}, F.~C., \& {Watson}, L.~C. 1998, \aj, 116, 2466

\bibitem[{{Gratton} {et~al.}(2012){Gratton}, {Carretta}, \&
  {Bragaglia}}]{gratton12}
{Gratton}, R.~G., {Carretta}, E., \& {Bragaglia}, A. 2012, \aapr, 20, 50

\bibitem[{{Gratton} {et~al.}(1996){Gratton}, {Carretta}, \&
  {Castelli}}]{gratton96}
{Gratton}, R.~G., {Carretta}, E., \& {Castelli}, F. 1996, \aap, 314, 191

\bibitem[{{Gratton} {et~al.}(2000){Gratton}, {Sneden}, {Carretta}, \&
  {Bragaglia}}]{gratton00}
{Gratton}, R.~G., {Sneden}, C., {Carretta}, E., \& {Bragaglia}, A. 2000, \aap,
  354, 169

\bibitem[{{Harris}(1996)}]{harris96}
{Harris}, W.~E. 1996, \aj, 112, 1487

\bibitem[{{Hurley} {et~al.}(2002){Hurley}, {Tout}, \& {Pols}}]{hurley02}
{Hurley}, J.~R., {Tout}, C.~A., \& {Pols}, O.~R. 2002, \mnras, 329, 897

\bibitem[{{Iben}(1967)}]{iben67}
{Iben}, Jr., I. 1967, \apj, 147, 624

\bibitem[{{Karakas}(2010)}]{karakas10}
{Karakas}, A.~I. 2010, in Principles and Perspectives in Cosmochemistry, ed.
  A.~{Goswami} \& B.~E. {Reddy}, 107

\bibitem[{{Kirby} {et~al.}(2012){Kirby}, {Fu}, {Guhathakurta}, \&
  {Deng}}]{kirby12}
{Kirby}, E.~N., {Fu}, X., {Guhathakurta}, P., \& {Deng}, L. 2012, \apjl, 752,
  L16

\bibitem[{{Kraft} {et~al.}(1999){Kraft}, {Peterson}, {Guhathakurta}, {Sneden},
  {Fulbright}, \& {Langer}}]{kraft99}
{Kraft}, R.~P., {Peterson}, R.~C., {Guhathakurta}, P., {Sneden}, C.,
  {Fulbright}, J.~P., \& {Langer}, G.~E. 1999, \apjl, 518, L53

\bibitem[{{Kumar} {et~al.}(2011){Kumar}, {Reddy}, \& {Lambert}}]{kumar11}
{Kumar}, Y.~B., {Reddy}, B.~E., \& {Lambert}, D.~L. 2011, \apjl, 730, L12

\bibitem[{{Lambert} {et~al.}(1980){Lambert}, {Dominy}, \&
  {Sivertsen}}]{lambert80}
{Lambert}, D.~L., {Dominy}, J.~F., \& {Sivertsen}, S. 1980, \apj, 235, 114

\bibitem[{{Lattanzio} {et~al.}(2015){Lattanzio}, {Siess}, {Church}, {Angelou},
  {Stancliffe}, {Doherty}, {Stephen}, \& {Campbell}}]{lattanzio15}
{Lattanzio}, J.~C., {Siess}, L., {Church}, R.~P., {Angelou}, G., {Stancliffe},
  R.~J., {Doherty}, C.~L., {Stephen}, T., \& {Campbell}, S.~W. 2015, \mnras,
  446, 2673

\bibitem[{{Lebzelter} {et~al.}(2012){Lebzelter}, {Uttenthaler}, {Busso},
  {Schultheis}, \& {Aringer}}]{lebzelter12}
{Lebzelter}, T., {Uttenthaler}, S., {Busso}, M., {Schultheis}, M., \&
  {Aringer}, B. 2012, \aap, 538, A36

\bibitem[{{Lind} {et~al.}(2009){Lind}, {Asplund}, \& {Barklem}}]{lind09}
{Lind}, K., {Asplund}, M., \& {Barklem}, P.~S. 2009, \aap, 503, 541

\bibitem[{{Martell} \& {Shetrone}(2013)}]{ms13}
{Martell}, S.~L., \& {Shetrone}, M.~D. 2013, \mnras, 624

\bibitem[{{Martin} {et~al.}(1994){Martin}, {Rebolo}, {Casares}, \&
  {Charles}}]{martin94}
{Martin}, E.~L., {Rebolo}, R., {Casares}, J., \& {Charles}, P.~A. 1994, \apj,
  435, 791

\bibitem[{{Momany} {et~al.}(2004){Momany}, {Bedin}, {Cassisi}, {Piotto},
  {Ortolani}, {Recio-Blanco}, {De Angeli}, \& {Castelli}}]{momany04}
{Momany}, Y., {Bedin}, L.~R., {Cassisi}, S., {Piotto}, G., {Ortolani}, S.,
  {Recio-Blanco}, A., {De Angeli}, F., \& {Castelli}, F. 2004, \aap, 420, 605

\bibitem[{{Monaco} {et~al.}(2011){Monaco}, {Villanova}, {Moni Bidin},
  {Carraro}, {Geisler}, {Bonifacio}, {Gonzalez}, {Zoccali}, \&
  {Jilkova}}]{monaco11}
{Monaco}, L., {et~al.} 2011, \aap, 529, A90

\bibitem[{{Monaco} {et~al.}(2014){Monaco}, {Boffin}, {Bonifacio}, {Villanova},
  {Carraro}, {Caffau}, {Steffen}, {Ahumada}, {Beletsky}, \&
  {Beccari}}]{monaco14}
---. 2014, \aap, 564, L6

\bibitem[{{Nataf} {et~al.}(2013){Nataf}, {Gould}, {Pinsonneault}, \&
  {Udalski}}]{nataf13}
{Nataf}, D.~M., {Gould}, A.~P., {Pinsonneault}, M.~H., \& {Udalski}, A. 2013,
  \apj, 766, 77

\bibitem[{{Pasquini} {et~al.}(2002){Pasquini}, {Avila}, {Blecha}, {Cacciari},
  {Cayatte}, {Colless}, {Damiani}, {de Propris}, {Dekker}, {di Marcantonio},
  {Farrell}, {Gillingham}, {Guinouard}, {Hammer}, {Kaufer}, {Hill}, {Marteaud},
  {Modigliani}, {Mulas}, {North}, {Popovic}, {Rossetti}, {Royer}, {Santin},
  {Schmutzer}, {Simond}, {Vola}, {Waller}, \& {Zoccali}}]{pasquini02}
{Pasquini}, L., {et~al.} 2002, The Messenger, 110, 1

\bibitem[{{Pilachowski} {et~al.}(2000){Pilachowski}, {Sneden}, {Kraft},
  {Harmer}, \& {Willmarth}}]{pilachowski00}
{Pilachowski}, C.~A., {Sneden}, C., {Kraft}, R.~P., {Harmer}, D., \&
  {Willmarth}, D. 2000, \aj, 119, 2895

\bibitem[{{Piotto} {et~al.}(2012){Piotto}, {Milone}, {Anderson}, {Bedin},
  {Bellini}, {Cassisi}, {Marino}, {Aparicio}, \& {Nascimbeni}}]{piotto12}
{Piotto}, G., {et~al.} 2012, \apj, 760, 39

\bibitem[{{Sackmann} \& {Boothroyd}(1999)}]{sb99}
{Sackmann}, I.-J., \& {Boothroyd}, A.~I. 1999, \apj, 510, 217

\bibitem[{{Siess} \& {Livio}(1999)}]{sl99}
{Siess}, L., \& {Livio}, M. 1999, \mnras, 308, 1133

\bibitem[{{Silva Aguirre} {et~al.}(2008){Silva Aguirre}, {Catelan}, {Weiss}, \&
  {Valcarce}}]{silva08}
{Silva Aguirre}, V., {Catelan}, M., {Weiss}, A., \& {Valcarce}, A.~A.~R. 2008,
  \aap, 489, 1201

\bibitem[{{Silva Aguirre} {et~al.}(2014){Silva Aguirre}, {Ruchti}, {Hekker},
  {Cassisi}, {Christensen-Dalsgaard}, {Datta}, {Jendreieck}, {Jessen-Hansen},
  {Mazumdar}, {Mosser}, {Stello}, {Beck}, \& {de Ridder}}]{silva14}
{Silva Aguirre}, V., {et~al.} 2014, \apjl, 784, L16

\bibitem[{{Skrutskie} {et~al.}(2006){Skrutskie}, {Cutri}, {Stiening},
  {Weinberg}, {Schneider}, {Carpenter}, {Beichman}, {Capps}, {Chester},
  {Elias}, {Huchra}, {Liebert}, {Lonsdale}, {Monet}, {Price}, {Seitzer},
  {Jarrett}, {Kirkpatrick}, {Gizis}, {Howard}, {Evans}, {Fowler}, {Fullmer},
  {Hurt}, {Light}, {Kopan}, {Marsh}, {McCallon}, {Tam}, {Van Dyk}, \&
  {Wheelock}}]{skru06}
{Skrutskie}, M.~F., {et~al.} 2006, \aj, 131, 1163

\bibitem[{{Smith} {et~al.}(1999){Smith}, {Shetrone}, \& {Keane}}]{smith99}
{Smith}, V.~V., {Shetrone}, M.~D., \& {Keane}, M.~J. 1999, \apjl, 516, L73

\bibitem[{{Sneden}(1973)}]{sneden73}
{Sneden}, C.~A. 1973, PhD thesis, THE UNIVERSITY OF TEXAS AT AUSTIN.

\bibitem[{{Zhang} {et~al.}(2015){Zhang}, {Kirby}, \& {Guhathakurta}}]{zhang15}
{Zhang}, A.~J., {Kirby}, E.~N., \& {Guhathakurta}, P. 2015, in American
  Astronomical Society Meeting Abstracts, Vol. 225, 449

\end{thebibliography}

\end{document}